\documentclass[useAMS,usenatbib]{mn2e}

\usepackage{graphicx}
\usepackage{xcolor}
\usepackage{amssymb, amsmath}
\usepackage{hyperref}
\usepackage[export]{adjustbox}
\hypersetup{colorlinks=true,citecolor=blue}
\usepackage[utf8]{inputenc}
\usepackage{amsmath, amssymb, bm}
\usepackage{multirow}
\usepackage{widetext}

\renewcommand{\vec}[1]{\bmath{#1}}
\newcommand{\Msun}{\ensuremath{M_{\odot}}}

\newcommand{\be}{\begin{equation}}
\newcommand{\ee}{\end{equation}}

\newcommand{\bea}{\begin{eqnarray}}
\newcommand{\eea}{\end{eqnarray}}
\newcommand{\bdm}{\begin{displaymath}}
\newcommand{\edm}{\end{displaymath}}

\newcommand{\eq}[1]{Eq.~(\ref{#1})}

\newcommand{\fig}[1]{Figure~\ref{#1}}

\def\ie{{\em i.e.}~}

\def\kMpc{\, h \, {\rm Mpc}^{-1}}

\title[Bias and shot-noise of the HI Power Spectrum]{On the spatial distribution of neutral hydrogen in the Universe: bias and shot-noise of the HI Power Spectrum}

\author[E. Castorina, F. Villaescusa-Navarro]
{Emanuele Castorina$^{1,2}$\thanks{e-mail: ecastorina@berkeley.edu}, 
Francisco Villaescusa-Navarro$^{3,4}$\thanks{e-mail: villaescusa@oats.inaf.it}
 \\~\\
\footnotesize
\footnotesize
$^1$Berkeley Center for Cosmological Physics, University of California, Berkeley, CA 94720, USA\\
$^2$Lawrence Berkeley National Laboratory, 1 Cyclotron Road, Berkeley, CA 93720, USA\\
$^3$INAF, Osservatorio Astronomico di Trieste, via Tiepolo 11, I-34131 Trieste, Italy\\
$^4$INFN -- National Institute for Nuclear Physics, Via Valerio 2, I-34127 Trieste, Italy\\
}

\begin{document}
\maketitle 

\begin{abstract}
The spatial distribution of neutral hydrogen (HI) in the Universe contains a wealth of cosmological information.  The 21 cm emission line can be used to map the HI up to very high redshift and therefore reveal us something  about the evolution of the large scale structures in the Universe. 
However little is known about the abundance and clustering properties of the HI over cosmic time.
Motivated by this, we build an analytic framework where the relevant parameters that govern how the HI is distributed among dark matter halos can be fixed using observations. At the same time we provide tools to study the column density distribution function of the HI absorbers together with their clustering properties.
Our formalism is the first one able to account for all observations at a single redshift, $z = 2.3$. 
The linear bias of the HI and the mean number density of HI sources, two main ingredients in the calculation of the signal-to-noise ratio of a cosmological survey, are then discussed in detail, also extrapolating the results to low and high redshift.
We find that HI bias is relatively higher than the value reported in similar studies, but the shot noise level is always sub dominant, making the HI Power Spectrum always a high signal-to-noise measurements up to $z\simeq5$ in the limit of no instrumental noise and foreground contamination.
\end{abstract} 
 

\section{Introduction}
\label{sec:introduction}

Our current understanding of the energy content of the Universe involve the presence of different components such as dark energy, dark matter, baryons or massive neutrinos. The interplay of the different constituents shape the spatial distribution of matter in the Universe. Information on the fraction that each component contributes to the overall energy budget of the Universe, together with information on the geometry and nature of the Universe initial conditions is thus embedded into the spatial distribution of matter. 

A way to constraint the value of the cosmological parameters is thus to measure the statistical properties of the distribution of matter in the Universe and compare them against the predictions of theoretical models. The problem resides in the fact that the distribution of matter is not directly observable. Our knowledge on it depends on the spatial distribution of tracers of it, such as galaxies, X-rays or cosmic neutral hydrogen (HI). 

In all cases the idea is that, on large-scales, the clustering properties of matter tracers, where perturbations are small, should resemble those of the underlying matter perturbations, modulo an overall normalization factor, which usually goes under the name of bias. Galaxy surveys such as the Sloan Digital Sky Survey\footnote{https://www.sdss3.org/surveys/boss.php} (SDSS) has mapped large regions of the sky at low-redshift and galaxy clustering measurements has been used to place tight constraints on the value of the cosmological parameters \citep[e.g.][]{Gil-Marin_2015, Alam_2016,Zhao_2016,Florien_2016a,Florien_2016b,Sanchez_2016,Grieb_2016}. 

The cosmic web can also be mapped with neutral hydrogen, which can be detected in the Universe either in absorption or in emission. In absorption it can be detected through the Ly$\alpha$-forest: the light from distant quasars can be absorbed by cosmic neutral hydrogen that it is located on its line of sight, producing a clear signature in their spectra. For instance, the clustering properties of the Ly$\alpha$-forest has been recently used to detect the BAO peak at $z=2.34$ \citep{Delubac_2015}. 

Cosmic neutral hydrogen can also be detected in emission through spectral lines such as the ${\rm H}_\alpha$ or the 21cm. In this paper we focus on the latter. The interaction between the spins of the electron and the proton induce a splitting on the hydrogen atom ground state; this is called the hyperfine structure. The wavelength of the energy difference between these two states is 21cm, while is frequency is 1420 MHz. In the post-reionization epoch, the typical temperatures of neutral hydrogen clouds range from tens to hundreds of Kelvin, much larger than the temperature different between the 2 hyperfine states, but smaller than the temperature required to excite the Ly$\alpha$ transition. Thus, in the post-reionization era neutral hydrogen clouds will have 3 out of 4 electrons in the hyperfine structure excited state, and their decay to the ground state will induce emission in terms of 21cm radiation. 

The 21cm emission by cosmic neutral hydrogen can be detected by radio-telescopes employing two different techniques. The first one is to detect directly the HI in galaxies (or in HI blobs \citep{Martin_2012,Villaescusa-Navarro_2016b, Burkhart_2016, Taylor_2016}); this is called a HI galaxy survey \citep{Yahya_2015, Abdalla_2015}. The second technique consists in carrying out intensity mapping observations, \ie performing low angular resolution radio-observation to measure the integrated 21cm radiation from large patches of the sky containing many galaxies, without resolving them individually \citep{Bharadwaj_2001A, Bharadwaj_2001B, Battye:2004re,McQuinn_2006, Chang_2008, Loeb_Wyithe_2008,Seo2010,Villaescusa-Navarro_2014a, Bull_2015,Santos_2015}. 

Each technique has its pros and cons. While a HI galaxy survey provides a catalogue with the location of galaxies (or HI clouds) from where we know very well how to extract the relevant cosmological information, the weakness of the signal would require of very powerful instruments, like the phase 2 of the Square Kilometre Array (SKA)\footnote{\url{https://www.skatelescope.org/}}, to be competitive with other surveys \citep{Bull_2015, Yahya_2015}. On the other hand, intensity mapping can be used to trace extremely large cosmological volumes but the theoretical framework requires more development and its complications (such as calibration, presence of large foregrounds, instrumental effects...etc) need to be fully understood and under control.  

In this paper we focus our attention on the 21cm intensity mapping technique, that given its spectroscopic nature, the large volumes it can sample and the isolation of the 21cm line \citep{Gong_2011} could revolutionize the field of cosmological observations. Current, upcoming and future instruments such as CHIME\footnote{\url{http://chime.phas.ubc.ca/}}, BINGO\footnote{\url{http://www.jb.man.ac.uk/research/BINGO/}}, ORT\footnote{\url{http://rac.ncra.tifr.res.in/}}, FAST\footnote{\url{http://fast.bao.ac.cn/en/}}, MeerKAT\footnote{\url{http://www.ska.ac.za/gallery/meerkat/}} or SKA1-MID will employ this technique to trace the large-scale structure of the Universe. In order to extract the maximum information from these surveys accurate theoretical models that model the observations are needed. 

Accurate predictions of the shape and amplitude of the fully non-linear 21cm power spectrum can be obtained using tools from the halo model \citep{Seljak_2000, PeacockSmith00,Scoccimarro2001, Cooray_2002}. The ingredients required are: the linear matter power spectrum, the halo mass function and halo bias, the relation between halo mass and HI mass (parametrized through the function $M_{\rm HI}(M,z)$) and the HI density profile within halos. On large, linear, scales, the HI density profile becomes irrelevant and the amplitude and shape of the signal is completely specified by the HI density parameter, $\Omega_{\rm HI}$, and by the bias of the neutral hydrogen tracer, $b_{\rm HI}$. Even if the HI field is mapped in a continuous way, the signal is coming from discrete sources, therefore estimates of the signal-to-noise from 21cm intensity mapping experiments also requires knowledge of the the effective number of tracers, in Power Spectrum analyses usually parametrized as a shot-noise term, $P_{\rm SN} = \bar{n}^{-1}$.

The cosmological abundance of neutral hydrogen is known at $1.5<z<5$, from measurements of the column density distribution function (CDDF) \citep{Noterdaeme12,Zafar13,Crighton15} and at lower redshifts from the HI mass function from surveys such as HIPASS\footnote{\url{http://www.atnf.csiro.au/research/multibeam/release/}} and ALFALFA\footnote{\url{http://egg.astro.cornell.edu/index.php/}} \citep{Zwaan_2005, Braun_2012}, and within the (large) errorbars it is very slowly evolving with redshift \citep{Crighton15}. 

Clustering properties of HI are instead completely unknown at all redshifts\footnote{We notice that what has been measured is the clustering of HI selected galaxies, not the clustering of HI itself \citep{Martin_2012}.}, although the product $\Omega_{\rm HI}b_{\rm HI}$ has been derived from intensity mapping observations at $z\simeq0.8$ by \cite{Switzer13}. This means that we do not know which halos host which amount of neutral hydrogen at a given redshift. Hydrodynamic simulations \citep{Bird14,Villaescusa-Navarro_2014a, Rahmati15,Villaescusa-Navarro_2016b} provide some insights but their results are not conclusive, since it is hard to account with existing models for different observations, and studying the abundance and spatial distribution of HI in numerical simulations requires hydrodynamic simulations with state-of-the-art feedback models coupled with radiative transfer calculations.
\begin{figure}
 \includegraphics[width=.45\textwidth]{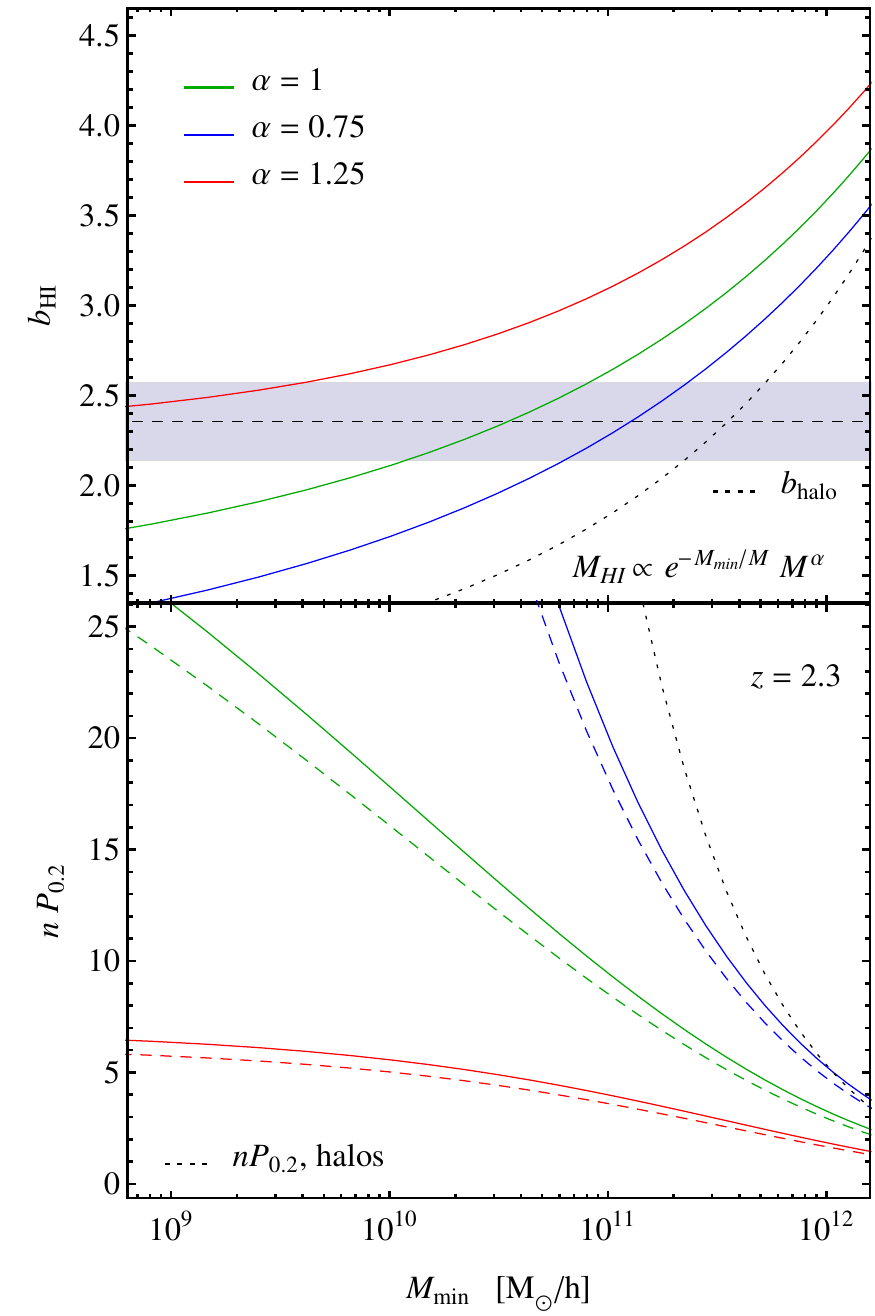}
 \caption{Top panel: linear HI bias factor as a function of the cutoff mass $M_{\rm min}$ for different values of $\alpha$. The bias of the corresponding dark matter halos is shown as a dotted line. Bottom panel: signal-to-noise ratio at $k=0.2 \kMpc$, continuous line for perfect BAO reconstruction and dashed line for 50\% reconstruction.}
 \label{fig:bHI}
\end{figure}
The closest one could get to a measurements of HI bias is the measurement from the BOSS collaboration\citep{Font-Ribera12} of the cross-correlation between the Lyman-$\alpha$ forest and the Damped-Lyman-$\alpha$ Systems (DLAs), which contain around 90\% of the all neutral hydrogen in the Universe. The quoted number for DLA bias is $b_{\rm DLA} = (2.17\pm0.2) \beta_F^{0.22}$, where $\beta_F$ is a number of order 1.5 \citep{CieplakSlosar15,Prats15}. However, as we shall see later, there is a crucial difference between the HI bias and the DLA bias, with non trivial observational consequences.

The goal of this paper is to present a new analytic formalism to model the spatial distribution of cosmic neutral hydrogen in the post-reionization era that can reproduce the observations.

Since the redshift evolution of the HI clustering is not constrained at all, we will fix the free parameters of our model at the single redshift where more data are available, $z_{ref}=2.3$. The measurements employed are therefore the abundance of Lyman Limit System (LLS) and DLAs from from \cite{Zafar13,Noterdaeme12} together with $\Omega_{\rm HI}$, and the DLA bias of \citep{Font-Ribera12}.
We then investigate the implications of our model, in terms of the bias and the shot-noise of the HI Power Spectrum, also including estimates of the signal-to-noise ratio. Since we are interested in the cosmological signal current or future 21 cm surveys could in principle measure, as a proof of concept we do not include instrumental systematics errors as well as foregrounds contamination in our calculations.

This paper is organized as follows.  In Sec \ref{sec:Form} our analytical model is introduced and we define all the relevant HI quantities in a halo model fashion. Then in \ref{sub:dsigma} we describe in a novel way how to make contact with the observations and fix the free parameters of our model in a consistent manner. Sec. \ref{sec:Res} contains a number of results relevant for observations of the HI Power Spectrum at our reference redshift. Finally in Sec \ref{sec:bOfz} we speculate on redshift evolution of HI properties within our model, to both low and high redshift. We draw the main conclusions of this work in section \ref{sec:conclusions}.
 
The cosmology used in this paper is a baseline $\Lambda$CDM cosmology from Planck 2015 \citep{Planck15}.

\begin{figure}
 \includegraphics[width=.45\textwidth]{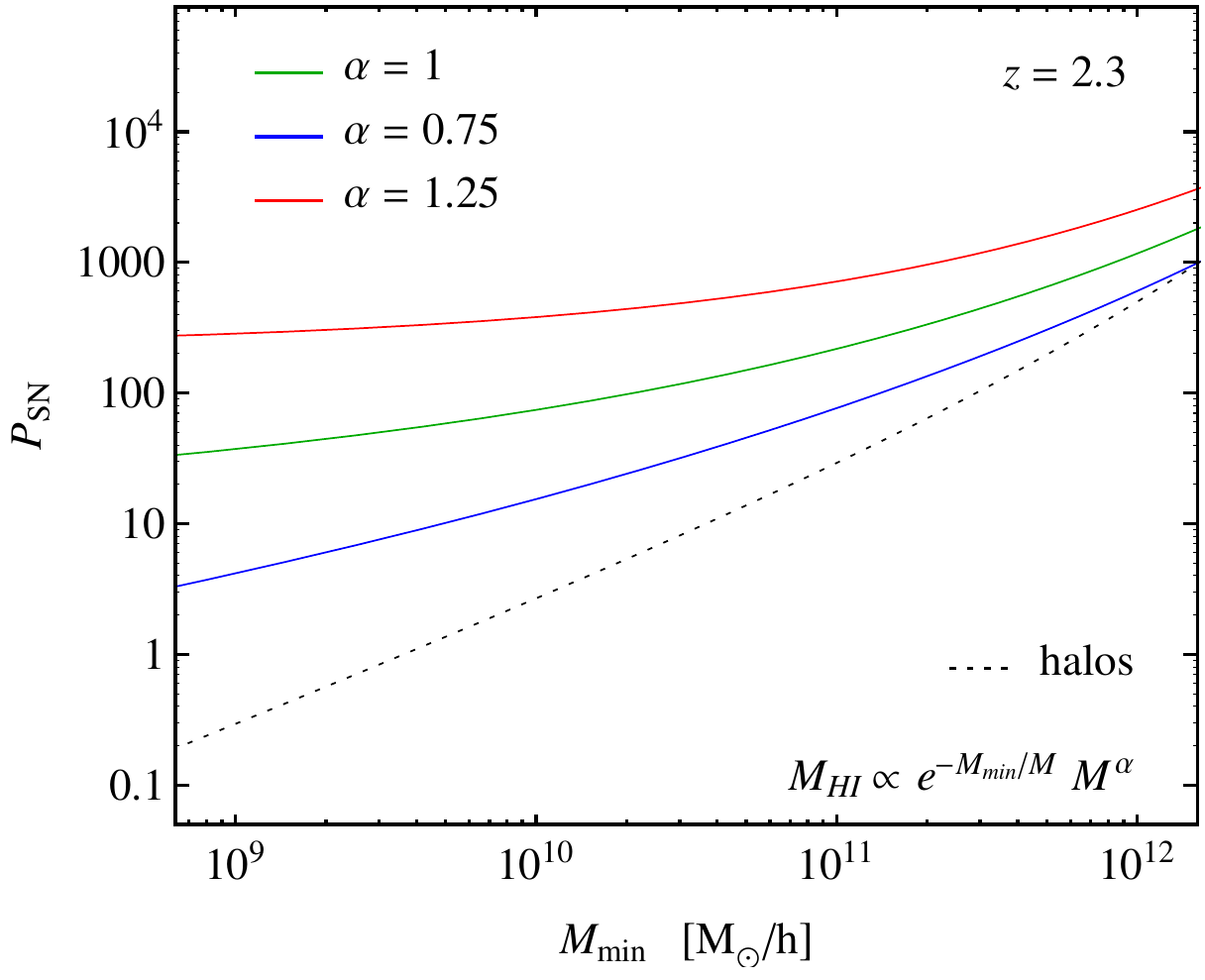}
 \caption{Shot noise contribution to the HI Power Spectrum for the model specified by \eq{eq:MHI} as a function of the cutoff mass $M_{min}$ and for different values of $\alpha$. The shot noise for the corresponding halo population is shown as a dotted line.}
 \label{fig:SN}
\end{figure}
\section{Formalism}
\subsection{Neutral hydrogen distribution in the halo model}
\label{sec:Form}
In the framework of the halo model \citep{Seljak2000,PeacockSmith00,Scoccimarro2001}, and therefore assuming that the contribution of HI outside halos is negligible \citep{Villaescusa-Navarro_2014a}, the cosmological abundance of neutral hydrogen can be written as
\be
\Omega_{\rm HI}(z)=\frac{1}{\rho_c^0}\int_0^\infty n(M,z)M_{\rm HI}(M,z)dM
\label{eq:Omega_HI}
\ee
where $n(M,z)$ is the halo mass function at redshift $z$, $M_{\rm HI}(M,z)$ is the average HI mass that a dark matter halo of mass $M$ contains at redshift $z$ and $\rho_c^0$ is the current critical density of the Universe today. A convenient reformulation of the mass function and of the above formula is
\be
\nu f(\nu,z)=\frac{M^2}{\rho_m}n(M,z)\frac{d\ln M}{d\ln \nu}
\ee
and
\be
\Omega_{\rm HI}(z)=\Omega_m\int_0^\infty f(\nu) \frac{M_{\rm HI}(\nu)}{M(\nu)}d\nu
\ee
where $\rho_m=\Omega_m\rho_c^0$ and the peak height is defined as $\nu=\delta_c/\sigma(M)$, where $\delta_c=1.686$ and $\sigma^2(M)$ is the variance of the linear density field once it is smoothed with a top-hat filter of radius $R$. The theoretical halo mass function used in this work is the best fit to N-body simulations from \cite{Tinker08}.
Similarly for the shot noise contribution, \ie the $k \rightarrow 0$ limit of the $1$-halo term, one writes
\be
P_{\rm SN} = \frac{1}{\rho_c^0}\frac{\Omega_m}{\Omega_{\rm HI}^2(z)}\int f(\nu) \frac{M_{\rm HI}^2(\nu)}{M(\nu)} d \nu
\label{eq:PSN}
\ee
while for the bias of absorbers of a given column density $N_{\rm HI}$ and for the bias of HI we have
\be
\label{eq:bDLA}
b_{\rm N_{HI}} = \cfrac{{ \displaystyle \int_0^\infty} d\nu f(\nu) b(\nu) \cfrac{\sigma_{\rm N_{HI}}(\nu)}{M(\nu)}}{{ \displaystyle \int_0^\infty} d\nu f(\nu) \cfrac{\sigma_{\rm N_{HI}}(\nu)}{M(\nu)}} 
\ee
and
\be
\label{eq:bHI}
b_{\rm HI} = \frac{\Omega_m}{\Omega_{\rm HI}(z)} { \displaystyle \int_0^\infty}d\nu f(\nu) b(\nu) \cfrac{M_{\rm HI}(\nu)}{M(\nu)}
\ee
with $\sigma_{N_{\rm HI}}$ the absorber cross section, \ie the projected area of the halo occupied by objects with a given $N_{\rm HI}$, and $b(\nu)$ the halo bias. As for halo the mass function, we use the halo bias calibrated on N-body simulations from \cite{Tinker10}.
\begin{figure}
\includegraphics[width=.45\textwidth]{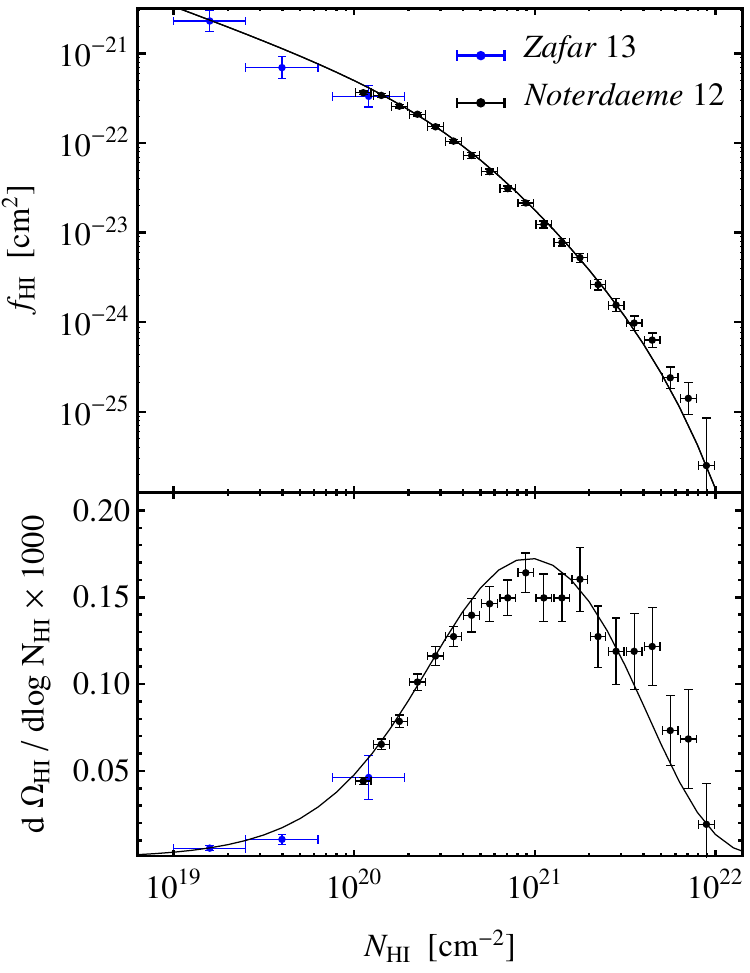}
\caption{Top panel: measurements of the CDDF at $z=2.3$ from \citep{Noterdaeme12,Zafar13}. The continuous line is the fit to the model described in \ref{sub:SI}. Bottom panel: the same measurements but expressed in terms of the differential contribution to $\Omega_{\rm HI}$, to reduce the dynamic range.}
 \label{fig:fHI}
\end{figure}
Without loss of generality we model the neutral hydrogen mass within halos as
\be
M_{\rm HI}(M,z) = \mathcal{C}(z) \,(1-Y_p)\, \frac{\Omega_b}{\Omega_m} e^{-M_{\rm min}(z)/M} \,M^{\alpha(z)}
\label{eq:MHI}
\ee
where $Y_p = 0.24$ is the Helium fraction, $M_{\rm min}$ represents a halo mass below which the HI abundance in halos is exponentially suppressed, and $\mathcal{C}$ is a normalization constant fixed using \eq{eq:Omega_HI}. 
This functional form is chosen so that the high-mass end of the $M_{\rm HI}(M)$ function follows a power-law, as found in hydrodynamic simulations and semi-analytic models 
\citep{Villaescusa-Navarro_2016a, Villaescusa-Navarro_2016b, Kim_2016} with the value of its slope controlling how efficient processes responsible for the creating and destruction of HI are. The exponential cutoff takes into account the fact that low mass halos are not expected to host a significant amount of HI since self-shielding becomes inefficient \citep{Pontzen_2008, Marin_2009, Villaescusa-Navarro_2014a}.
The power law index $\alpha$ regulates how fast is HI mass accreted onto halos. The same functional form has been adopted in previous studies by \cite{Barnes14,Padmanabhan15,Villaescusa-Navarro_2014a,Padmanabhan16a,Padmanabhan16b,Seehars2015} while \cite{Bagla2010} proposed a different formula which however poorly fits the bias of the DLAs \citep{Villaescusa-Navarro_2014a}. 
Hydrodynamic simulations \citep{Villaescusa-Navarro_2016a, Villaescusa-Navarro_2016b} and semi-analytic models \citep{Kim_2016} have a preference for $\alpha \le 1$, since processes such as tidal interactions, ram-pressure stripping and mergers, which tend to remove HI from the galaxies, appear to be more efficient than those that stimulate the cooling of the hot gas.

Equations \ref{eq:bDLA} and \ref{eq:bHI} show the main difference between the bias of the DLAs, objects with $N_{\rm HI} > 10^{20}~{\rm cm^{-2}}$, and the bias of neutral hydrogen\footnote{Most often in the literature DLAs are defined as objects with $N_{\rm HI} > 10^{20.3}~{\rm cm^{-2}}$. The definition in the main text is however the one employed by \cite{Font-Ribera12} to measure the DLA bias.}. While the former is number weighted, the latter is mass weighted, therefore, a-priori there is no reason to assume they are equal or very close to each other as commonly done in the literature. Already by dimensional analysis, and assuming a top-hat universal HI profile in halos, for a power law index $\alpha$ in \eq{eq:MHI}, we see that the cross section $\sigma_{\rm DLA}\propto \alpha^{2/3}$.

We conclude this section with some general consideration on HI bias and shot noise at $z=2.3$. The way the normalization of the HI-halo relation enters in Equations \ref{eq:bHI} and \ref{eq:PSN} makes $b_{\rm HI}$ and $P_{\rm SN}$ independent from the value of the cosmological abundance of neutral hydrogen $\Omega_{\rm HI}$. This means that those two quantities are functions of $M_{\rm min}$ and $\alpha$ only. 
The top panel of \fig{fig:bHI} shows predictions for he HI bias as a function of the cutoff mass $M_{\rm min}$ for three different values of $\alpha = 0.75,1,1.25$. The horizontal band is the mean value and $1\sigma$ error of DLAs bias from \cite{Font-Ribera12}.
As expected the HI bias increases as we remove hydrogen from low mass halos, and in general higher values of the power law index $\alpha$ translates into higher HI bias. We also note that $b_{\rm HI}$ is always larger than the bias of the halos, shown as a dotted line, above a given minimum mass.

The shot-noise term, \ie the inverse of the effective mean number density of HI sources, predicted by \eq{eq:PSN} is shown in \fig{fig:SN} for the same parameters of \fig{fig:bHI}. The mass weighting of the HI makes the shot noise larger than one of the same populations of halos, for the relatively low $M_{min}$ we are interested in this work, and, as expected, higher power laws in the $M_{\rm HI}(M)$ relation yield higher shot noise values.
For comparison, at $M_{\rm min} \simeq 10^{11}~h^{-1}M_\odot$, the HI number density is approximately 20 times larger than the number density of galaxies in BOSS at $z=0.57$, despite the fact that HI bias is comparable to the bias of luminous red galaxies (LRGs) in the CMASS sample \citep{BOSSDR12}.

As a first attempt to investigate the signal-to-noise of a 21cm experiment targeting the HI Power Spectrum via intensity mapping, we compute the following quantity, usually relevant for BAO studies,
\be
\label{eq:nP02}
n\,P_{0.2} = \frac{b_{\rm HI}^2 P_L (k=0.2 \,\kMpc, z=2.3)}{P_{\rm SN}}
\ee
where $P_{L}$ is the linear power spectrum and $P_{\rm SN}$ is computed using \eq{eq:PSN}. This estimate has to be intended for illustrative purposes only, since it does not include non-linear evolution, redshift space distortions, residual foreground contamination, angular resolution effects, beam uncertainties, foreground wedges and so on, which are pretty much experiment dependent \citep{Seo_2016,Cohn2016,Bigot_2015,Alonso_2015}.
On the other hand it quantifies the purely \emph{cosmological} signal present in the HI field.
In the bottom panel of \fig{fig:bHI} we plot the value of $nP_{0.2}$ for several values of $M_{\rm min}$ and $\alpha$. The continuous lines show \eq{eq:nP02}, \ie perfect BAO reconstruction \citep{Eisenstein07,Padmanabhan09}, whereas the dashed ones 50\% reconstruction, in the sense of \cite{Font-Ribera14}.

As we have seen before in Figures \ref{fig:bHI} and \ref{fig:SN} the HI bias and shot noise  increase with $\alpha$, hence the opposite trend is expected for the signal-to-noise. For $\alpha = 0.75,1$ the signal-to-noise is very high until $M_{\rm min} \simeq  4\times 10^{10} h^{-1}\Msun$, when shot noise starts to become more importat. For $\alpha = 1.25$ the number density of tracers is too low so $n\, P_{0.2}$ is always considerably smaller then the other two cases, but still larger than one up to $M \simeq 10^{12} \Msun$. 
We also note that since 21cm observations would mostly make observations at redshifts where non-linearities are less important, the gain introduced by BAO reconstruction is only marginal, which can be seen by comparing the continuous lines and the dashed ones in \fig{fig:bHI}.

The goal of the next section would be to provide a framework to fix $M_{\rm min}$ and $\alpha$ using the bias of the DLAs and measurements of the CDDF.
\subsection{A model for the cross section of $N_{\rm HI}$ systems}
\label{sub:dsigma}
Previous studies \citep{Barnes14,Padmanabhan15,Padmanabhan16a,Padmanabhan16b,Villaescusa-Navarro_2014a} have focused on modeling the HI profile within halos to compute absorbers cross section and distribution function. Common parametrizations are modified NFW profiles, with different concentration and/or slopes. However, not only this choice is completely unjustified \citep[see][for some observational studies at low redshift]{Wang_2014}, but those model suffer the problem of being incompatible with the DLA bias or yielding a too large $\Omega_{\rm HI}$ compared to the data.
\begin{figure}
 \includegraphics[width=.45\textwidth]{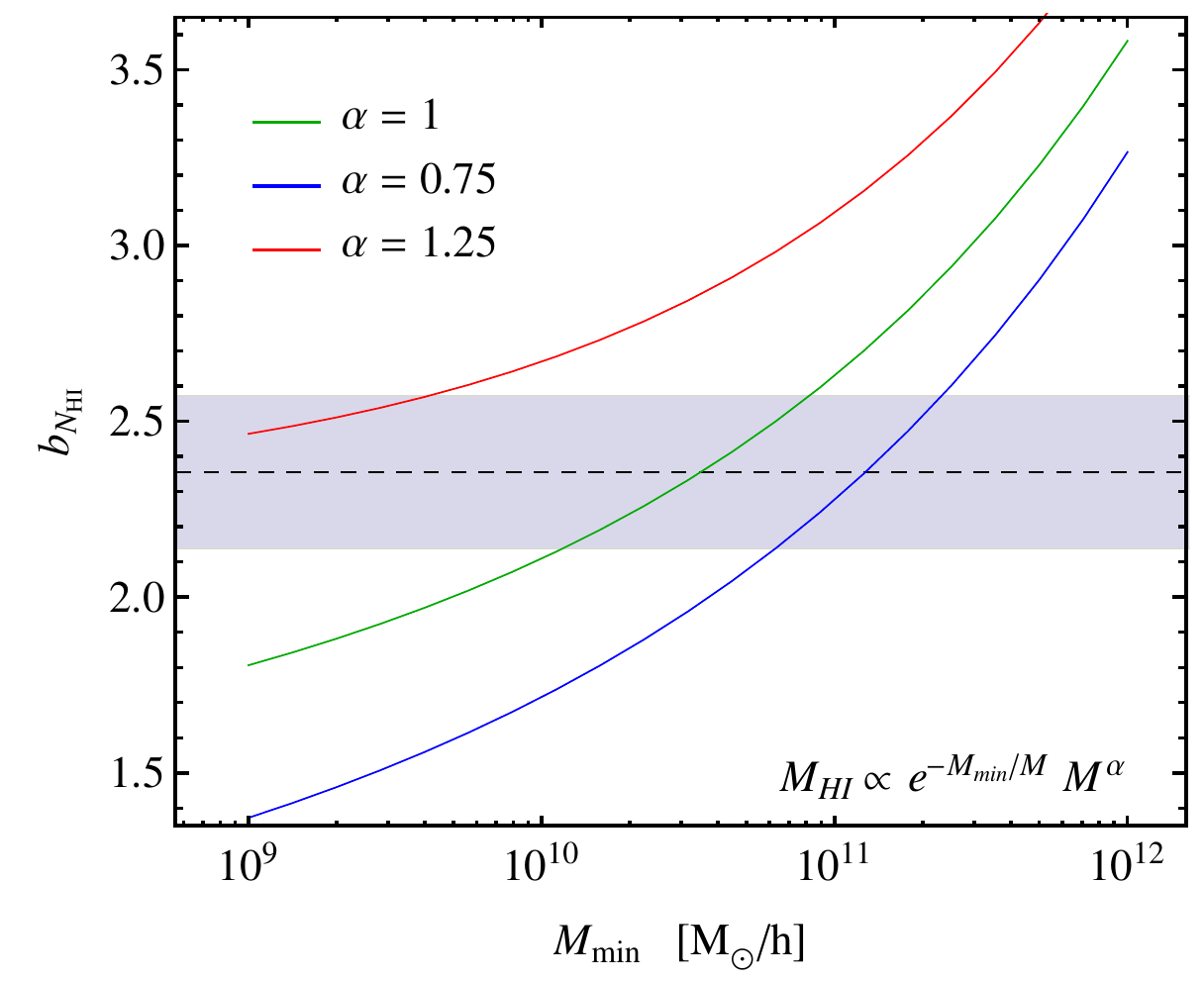}
 \caption{Bias of any $N_{\rm HI}$ absorbers as a function of $M_{\rm min}$ and for different $\alpha$s for the model described in \ref{sub:SI}}
 \label{fig:bNHI}
 \end{figure}
\begin{figure}
\includegraphics[width=.45\textwidth]{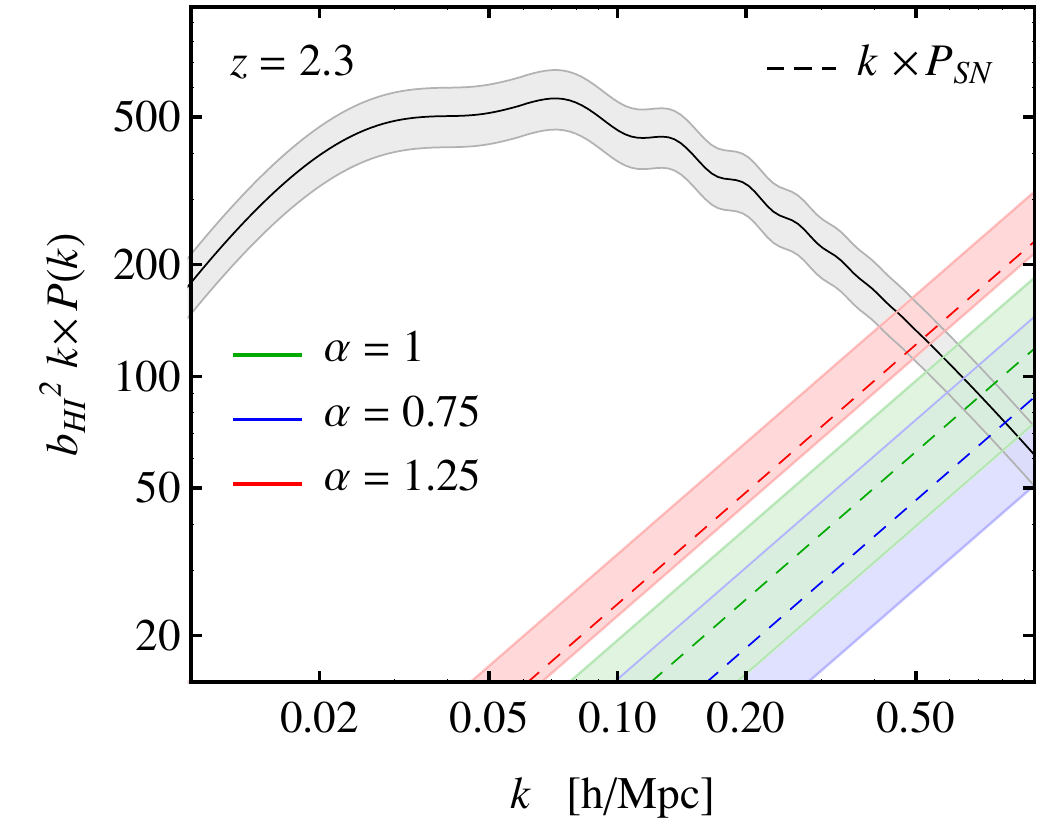}
\caption{Power spectrum of the HI at $z=2.3$. The bias is the same for the different values of $\alpha$, but the shot noise is different.}
\label{fig:PSNHI}
\end{figure}
Rather than modeling the HI density profile, we, for the first time, try model the cross section of absorbers directly. Suppose that in a halo of mass $M$, virial radius $R$ and total HI mass $M_{\rm HI}$, absorbers with a column densities in the range $[N_{\rm HI},N_{\rm HI}+dN_{\rm HI}]$ occupy an area $d\sigma$. We can write the following identity
\begin{equation}
m_{\rm H}\int_0^\infty \frac{d\sigma}{dN_{\rm HI}}(N_{\rm HI};M)N_{\rm HI}(M)dN_{\rm HI}=M_{\rm HI}(M)
\end{equation}
which arises just from the definition of $M_{\rm HI}$ 
\begin{eqnarray}
M_{\rm HI}(M,z)&=&\int_{V}d^3\vec{x}\rho_{\rm HI}(\vec{x};M)\\
&=&\int_V d^2\vec{r}dz\rho_{\rm HI}(\vec{r},z;M) \notag\\
&=&\int_Sd^2\vec{r}N_{\rm HI}(\vec{r};M,z)m_{\rm H} \notag \\
&=&\int_{N_{\rm HI}} dN_{\rm HI}\frac{d\sigma}{dN_{\rm HI}}(N_{\rm HI};M,z)N_{\rm HI}(M,z)m_{\rm H} \notag
\label{eq:MHI_rel}
\end{eqnarray}
where we moved from Cartesian to cylindrical coordinates (first to second line) and assumed that absorbers with column densities  in the range $[N_{\rm HI},N_{\rm HI}+dN_{\rm HI}]$ occupy an area $d\sigma$ (from third to fourth line). The hydrogen atom mass is $m_{\rm H}$. Notice that here we are not assuming anything about the spatial distribution of absorbers. It can be that absorbers within the virial radius do not occupy an area equal to that of the halo, so in general
\be
\int_0^{\infty}\frac{d\sigma}{dN_{\rm HI}}dN_{\rm HI}\neq\pi R^2
\ee
In this formalism, the DLAs cross-section is simply given by
\be
\sigma_{\rm DLA}(M)=\int_{10^{20}}^\infty \frac{d\sigma}{dN_{\rm HI}}(N_{\rm HI};M)dN_{\rm HI}
\ee
and the HI CDDF is
\be
f(N_{\rm HI})=\frac{c}{H_0}\int_0^\infty n(M,z)\frac{d\sigma}{dN_{\rm HI}}(N_{\rm HI};M)dM
\ee
The advantage of this formalism is that we can place a cutoff on the function $d\sigma/dN_{\rm HI}$ for $N_{\rm HI}<10^{19}~{\rm cm}^{-2}$, since the Ly$\alpha$ forest contains $\le 1\%$ of the total neutral hydrogen in the Universe \citep{Noterdaeme12,Zafar13}. Furthermore, in our way of writing the relation between the cross section and the total HI mass, we avoid modeling the spatial distribution of HI within halos, that hydrodynamic simulations have shown to be sparse and highly concentrated around DLAs and LLS \citep{Pontzen_2008,Bird_2014,Villaescusa-Navarro_2014a,Rahmati_2015}, 
and just attempt to model the cross-section of HI absorbers, a quantity more closely related to observations.

One can also rewrite the bias of DLAs and HI in terms of the differential cross section, and the net results is 


\begin{eqnarray}
b_{\rm HI}(z)&=&\frac{\int_0^\infty b(M,z)n(M,z)dM\int_{0}^{\infty} d\sigma N_{\rm HI}}{\int_0^\infty n(M,z)dM\int_{0}^{\infty} d\sigma N_{\rm HI}}\\\nonumber \\
b_{\rm DLA}(z)&=&\frac{\int_0^\infty b(M,z)n(M,z)dM\int_{10^{20}}^\infty d\sigma}{\int_0^\infty n(M,z)dM\int_{10^{20}}^\infty d\sigma}\;,
\end{eqnarray}
where $d\sigma\equiv dN_{\rm HI}\cfrac{d\sigma}{dN_{\rm HI}}(N_{\rm HI}|M,z)$. Those equations can also be expressed as a function of the observed CDDF:
\begin{eqnarray}
b_{\rm HI}(z)&=&\frac{c}{H_0}\frac{\int_0^\infty b(M,z)n(M,z)dM\int_{0}^{\infty} d\sigma N_{\rm HI}}{\int_0^\infty f(z,N_{\rm HI})N_{\rm HI}~dN_{\rm HI}}
\label{eq:bHIDLAa}\\
b_{\rm DLA}(z)&=&\frac{c}{H_0}\frac{\int_0^\infty b(M,z)n(M,z) dM\int_{10^{20}}^\infty d\sigma}{\int_{10^{20}}^\infty f(z,N_{\rm HI})~dN_{\rm HI}}
\label{eq:bHIbDLA}
\end{eqnarray}
The above equations explicitly show that the bias of the HI and the bias of the DLAs are different in general, as already discussed on section \ref{sec:Form}, with HI bias weighing by column density the area covered by the different absorbers. Note also that the denominator in both expressions can be estimated directly from observational data.
\section{Results}
\begin{figure}
 \includegraphics[width=.45\textwidth,center]{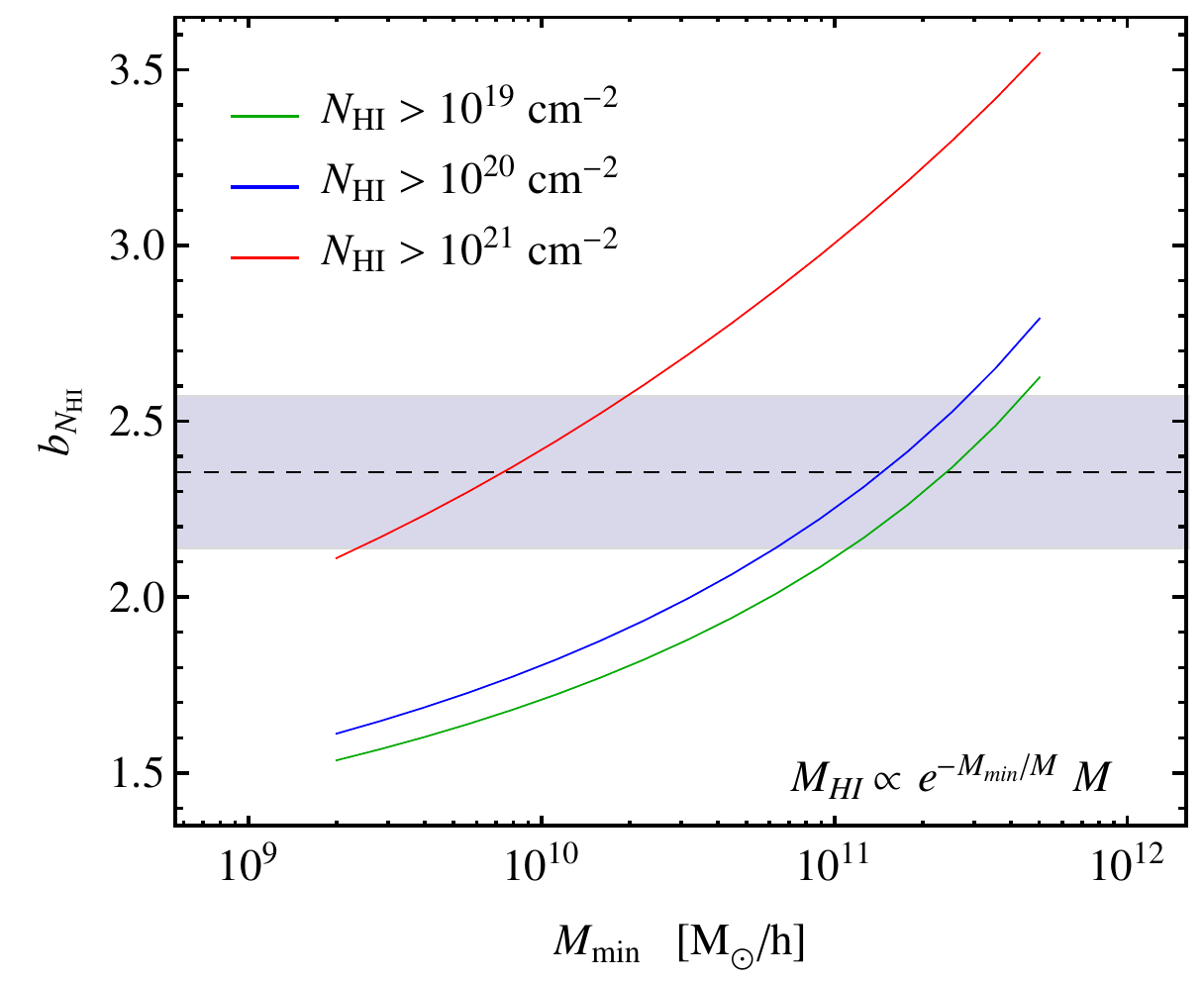}
 \caption{Bias of systems with different column densities for the model specified by \eq{eq:betaM} and $\alpha = 1$.} 
 \label{fig:bHIExp}
\end{figure}
\label{sec:Res}
Given the function $M_{\rm HI}(M,z)$ we just need to specify the differential cross section to have a complete description of the HI distribution.
The data of the CDDF at $z = 2.3$ \citep{Noterdaeme12,Zafar13}, see \fig{fig:fHI}, exhibits a low $N_{\rm HI}$  and a high $N_{\rm HI}$ cutoff with a power-law on intermediate values of the column density. Motivated by this, we choose the following form for the differential cross section,
\begin{equation}  
\frac{d\sigma}{dN_{\rm HI}}(N_{\rm HI}|M,z=2.3) = 
 A(M)N_{\rm HI}^{\beta}e^{-N_{\rm HI}/\gamma_1}(1-e^{-N_{\rm HI}/\gamma_2}) 
\label{eq:dsigma}
\end{equation} 
for $N_{\rm HI}\geqslant 10^{19}\,{\rm cm}^{-2}$, with $\beta$, $\gamma_1$ and $\gamma_2$ free parameters that could depend on both halo mass and redshift. 
The overall normalization, that depends on halo mass, $A(M)$ is chosen by requiring that the HI mass of all absorbers reproduce the total HI mass of the halo, i.e. by imposing Eq. \ref{eq:MHI}.
\begin{figure}
 \includegraphics[width=.45\textwidth,left]{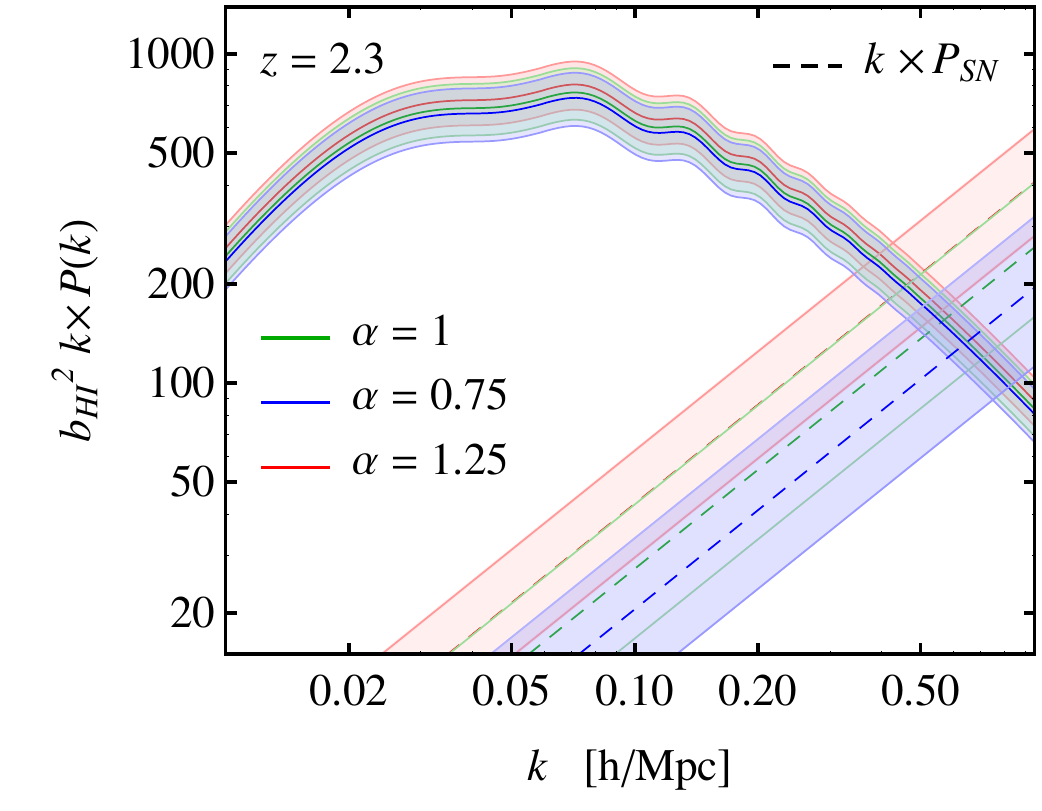}
 \caption{Same as in \fig{fig:PSNHI} but for the model specified by equation \eq{eq:betaM}.}
 \label{fig:SNExp}
\end{figure}
\subsection{Scale invariant model}
\label{sub:SI}
The case in which the three parameters of \eq{eq:dsigma} are independent of mass is special. It implies that the integrals in Equations \ref{eq:bHIDLAa} and  \ref{eq:bHIbDLA} are separable and therefore not only the bias of DLA and the bias of HI are the same, but the bias of all absorbers is the same. Physically, the fractional area covered by a given absorber with column density $N_{\rm HI}$ is the same regardless of the halo mass and $\alpha$. 

This result could be easily tested with data in the near future, simply splitting systems by column density and compare the bias of two different $N_{\rm HI}$ population. A preliminary study has been carried out by \cite{Font-Ribera12}, where no difference between different absorbers was found, although with large error bars.
The best fit parameters of this model are
\be
\label{eq:beta0}
\beta = -1.725\;,\; \gamma_1 = 10^{21.45} \,\rm{cm}^{-2}\;,\;\gamma_2 = 10^{20.32}\,\rm{cm}^{-2}
\ee
and in \fig{fig:bHI} the best fit model is compared against the data of the CDDF. 
The lower panel of \fig{fig:fHI} also shows that most of the contribution to the total HI in the Universe, and hence to HI bias, is coming from systems with $N_{\rm HI} \simeq 10^{21}~{\rm cm}^{-2}$ and their hosting halos.
The bias of $N_{\rm HI}$ systems, and of the HI in this model, is shown in \fig{fig:bNHI}, where one can read the cutoff mass that matches the observed value of $b_{\rm DLA}$. For the lower value of $\alpha$ favored by hydrodynamic simulations we need $M_{\rm min}\simeq 2 \times 10^{11} h^{-1}\Msun$.
With all the parameters fixed we can compute the shot noise contribution to the Power Spectrum which is not going to be the same  for different $\alpha$'s since $M_{\rm min}$ changes. This is shown in \fig{fig:PSNHI}. 
Since the error on DLA bias is still large, we show the $P(k)$ and the shot-noise as  bands, corresponding to the mean and $1-\sigma$ error of $b_{DLA}$. 
We conclude, together with \fig{fig:bHI}, that if HI absorbers are distributed within halos according to Equations (\ref{eq:dsigma}, \ref{eq:beta0}) the shot noise contribution to the Power Spectrum is always sub-dominant with respect to the cosmological signal.

\subsection{Mass dependent models}
\label{sub:Exp}
\begin{figure}
  \includegraphics[width=.45\textwidth]{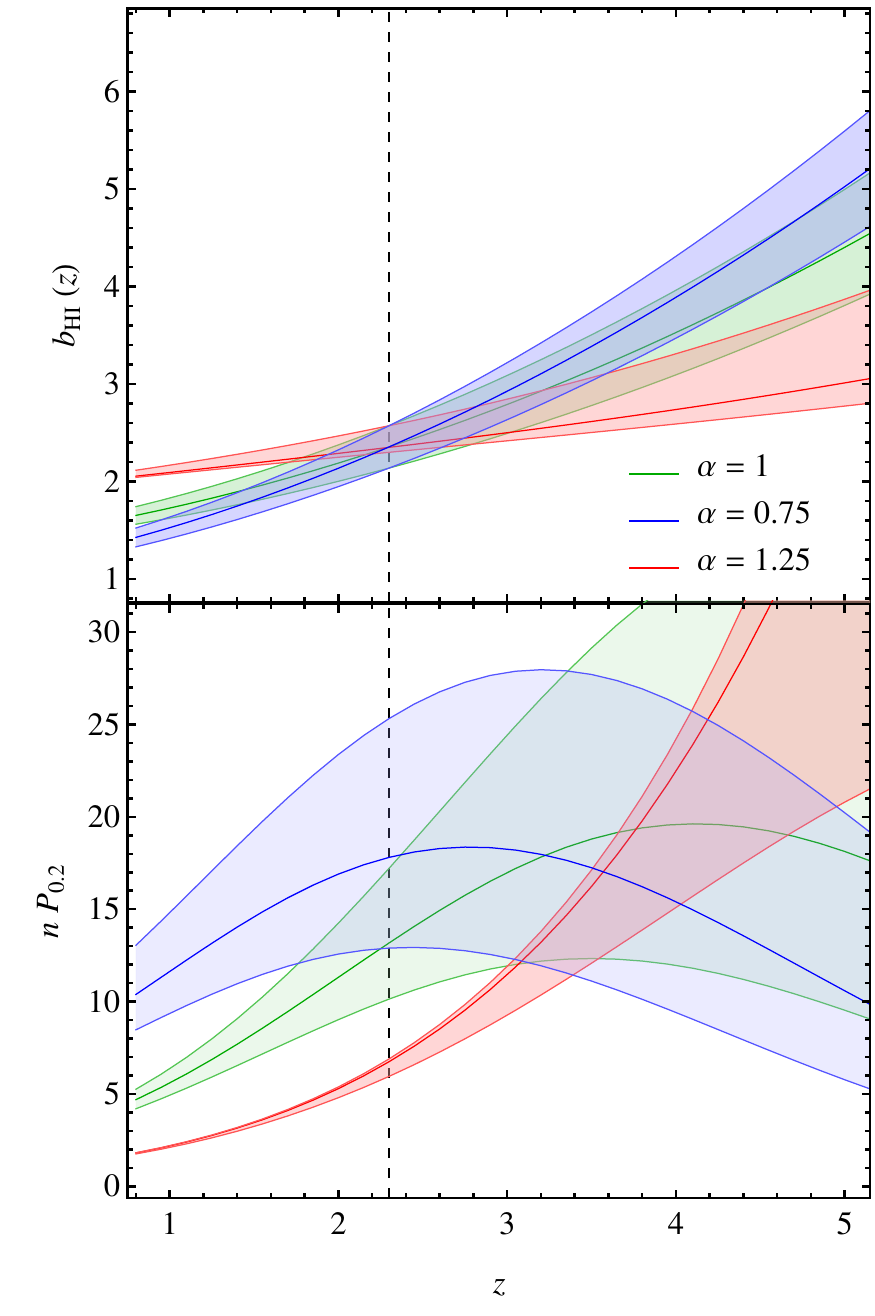}
 \caption{Evolution of the linear HI bias with redshift. The best fit parameters have been fixed using the mode described in \eq{eq:beta0} and we assume they are independent of redshift. See text for more discussion}
 \label{fig:bHIz}
\end{figure}
It is worth exploring models beyond the simple one described above. While scenarios with $b_{\rm HI}< b_{\rm DLA}$ are certainly plausible, their phenomenology is not very different from the description in Section \ref{sub:SI},  yielding similar values, actually even larger, for the signal-to-noise ratio at $k = 0.2 \kMpc$.

On the other hand, the HI bias can be arbitrary larger than $b_{\rm DLA}$, and therefore the HI $P(k)$ more severely affected by shot noise. We seek for one such a model in this section. 

In practice we want to remove high column density systems from low mass halos and move them to high mass halos. The easiest way to realize this is to make the slope $\beta$ in \eq{eq:dsigma} mass dependent, and we found that the following form gives very good fit to the measurements of the CDDF,
\be
\label{eq:betaM}
\beta(M) = \beta_0 e^{-M/M_0}
\ee
where $M_0=2.5\times10^{12}~h^{-1}\Msun$. Different choices for the cutoff mass in the above equation could be made, with the effect of changing the minimum mass we remove low $N_{\rm HI}$ objects from\footnote{We have checked that for different choices of the cutoff in \eq{eq:betaM} the final values of bias and shot noise are very similar.}.

High column density systems will now be more highly biased that the low column density ones. This is shown in \fig{fig:bHIExp} for $\alpha=1$, where we also notice that the cutoff mass needed to reproduce the DLA bias  has shifted from $M_{\rm min}\simeq 5 \times 10^{10}~h^{-1}M_\odot$ of \fig{fig:bNHI} to $M_{\rm min} \simeq 2\times 10^{11}~h^{-1}M_\odot$. As discussed in the previous section, if future measurements will point towards the same bias for different $N_{\rm HI}$ systems, this model would be excluded.

We repeated the analysis for the other two values of $\alpha$, finding the minimum mass that fits the bias of the DLAs, and in \fig{fig:SNExp} we summarize the results for the model described in this Section. The shot noise is higher than the case described in \ref{sub:SI}, and it reduces the value of $nP_{0.2}$ to
\be
n\,P_{0.2} \simeq  9,\,7\,,4   \quad \text{for}\quad \alpha = 0.75,1,1.25\;,
\ee
when $b_{DLA} = 2.4$.
Another important difference with the model described in the previous section is that HI bias will not be the same anymore for different values of $\alpha$, with higher $\alpha$'s predicting higher HI bias, see \fig{fig:SNExp}. Overall the decrement in signal-to-noise at $k=0.2 \kMpc$ between the two models is relatively large, around $50\%$ for low $\alpha$'s, and $40\%$ for $\alpha=1.25$. Nevertheless the value of $nP$ always safely remains  above $1$.
\label{sec:bOfz}
\begin{figure}
 \includegraphics[width=.45\textwidth]{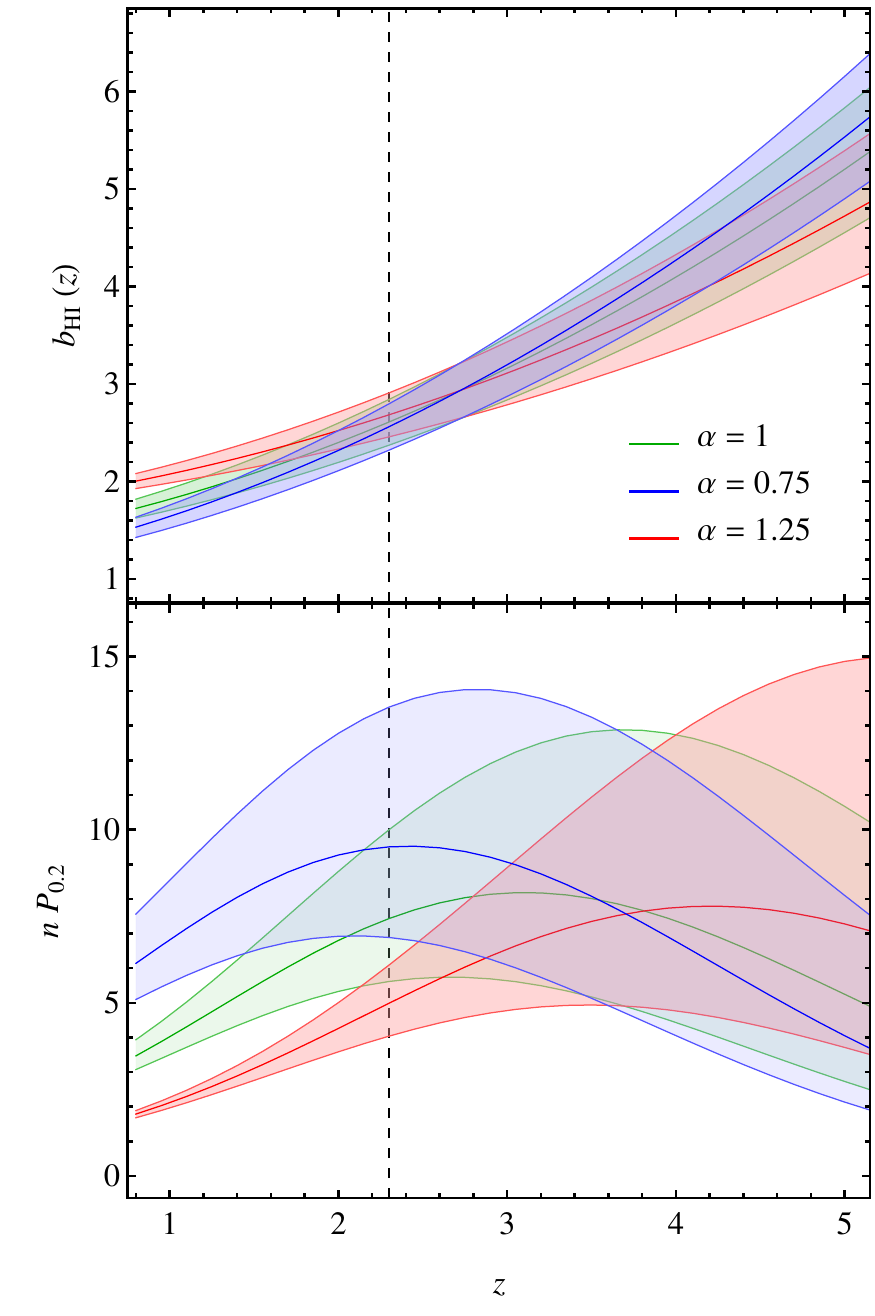}
 \caption{Same as \fig{fig:bHIz}  but using the best fit parameters in \eq{eq:betaM}.}
 \label{fig:bHIzExp}
\end{figure}
We notice, from \fig{fig:bHI}, that to considerably reduce the signal-to-noise in the HI $P(k)$ one would need $M_{min}\simeq 5\times 10^{12} \Msun$, right at the cutoff of in the halo mass function at $z=2.3$. The feasibility of such models is debatable, requiring a very strong UV background that keeps the gas ionized in very massive halos. We also find harder to fit the observations with such a large $M_{min}$. 

Models for the cross-section which yield $b_{\rm HI}<b_{\rm DLA}$ can be easily obtained by flipping the ratio in \eq{eq:betaM}. Their predictions go in the opposite direction of what we have shown in Figures \ref{fig:bHIExp} and \ref{fig:SNExp}, with an actual increase of signal-to-noise compared to the model in Section \ref{sub:SI}.

\section{Redshift evolution of HI clustering}
The redshift evolution of the HI bias and of the number density of HI sources would be of great importance for cosmological studies. Not only for low redshift 21cm surveys like CHIME, BINGO, SKA1-MID, but also for high redshifts ones that could measure the distribution of neutral hydrogen up to the end of reoinization. In particular SKA is expected to yield measurements of the HI field up to the end of reionization, $z\simeq6$, and the Cosmic Visions documents \citep{Visions16} have proposed a dedicated 21cm experiment above $z>3$ as a one of the five possible projects for the next generations of cosmological surveys.

It is therefore interesting to study how the HI field evolves across cosmic time within the framework described in this paper. 

However, the redshift evolution of the HI field is mostly unconstrained from data, with the exception of $\Omega_{HI}$ which is almost independent of $z$, making the dependence of $M_{\rm min}$ and $\alpha$ on redshift in principle arbitrary. In the simplest case it can be assumed that they do not depend on redshift, and the normalization constant in \eq{eq:MHI} is fixed at each redshift to match $\Omega_{\rm HI}(z)$. 
Hydrodynamic simulations by \cite{Villaescusa-Navarro_2016b} show that below $z\simeq2$ this could be the case, hence we adopt this choice, also extrapolating it to higher redshift, $z<5$.

It is important to notice that in repeating the calculation outlined in Sections \ref{sub:SI}-\ref{sub:Exp} at different redshifts, we do not need to specify the cosmological HI density, $\Omega_{\rm HI}(z)$, since HI bias and shot noise are independent of the overall normalization. 

This means that any model based on Equations \ref{eq:MHI}-\ref{eq:dsigma} could be further constrained by measurements of $\Omega_{\rm HI} b_{\rm HI}$, the main quantities that enters the $21$cm $P(k)$, combined with data on $\Omega_{\rm HI}$ from data of the CDDF or the HI mass function.

At $z=0.8$ measurement of the 21cm auto $P(k)$ from \citep{Switzer13} reported 
$\Omega_{\rm HI} b_{\rm HI} = [0.62 ^{+0.23}_{-0.15}] \,\times 10^{-3}$ which together with the abundance of HI at $z=1$ from \citep{Rao06} implies that $b_{\rm HI}(z\simeq0.8)\simeq1.5$, with however large error bars. 

In \fig{fig:bHIz}(\ref{fig:bHIzExp}) we show prediction of the redshift evolution of the bias and the shot noise for the models described in Section \ref{sub:SI}(\ref{sub:Exp}). We notice, as expected, a large decrease in the signal-to-noise at $k=0.2~h{\rm Mpc}^{-1}$ at all redshift in models with mass dependence on DLA cross section, notice the difference in the y-scale between \fig{fig:bHIzExp} and  \fig{fig:bHIz}. For $z<z_{ref} = 2.3$, the drop in the signal-to-noise ratio is due to the different redshift evolution of linear bias and shot noise, with the latter winning over the former for low $M_{min}$ and low $z$.
We therefore conclude that above $z=3$ the HI Power Spectrum remains a high signal-to-noise measurement. This statement applies for the models discussed in this paper, which are anyway fairly general, and it could be assessed in more details once new measurements are available. We warn again the reader that our estimates does not include instrumental or systematics uncertainties that usually become harder to treat at higher redshift.
Other approaches to redshift evolution of the bias, such as passive evolution \citep{Fry96}, or linear evolution models \citep{Reid14}, are not able to predict the evolution of the number density of tracers. Nevertheless they predict too large values of HI bias at low redshift, $b_{\rm HI}(z\simeq1)\simeq2$.

\section{Conclusions}
\label{sec:conclusions}

We have shown, for the first time, how to construct a consistent model for the distribution of the neutral hydrogen in the Universe, which can account for all the existing observations at $z_{ref} = 2.3$. 

Our model makes predictions for the bias and the shot noise of the HI Power Spectrum, and for the bias of HI absorbers with different column density. Those results could be tested with future data that would allow to further constraint the remaining freedom in the parameter space of the model.

We have then computed the signal-to-noise ratio of the HI Power Spectrum at $k=0.2 \kMpc$, a scale relevant for BAO studies, and concluded that, apart from instrumental and foreground systematics, the distribution of neutral hydrogen can be a very powerful cosmological observable.
Lacking observations, and therefore robust modeling, of the redshift evolution of the HI, we have made conservative assumptions for the evolution of $b_{HI}$ and $P_{SN}$, drawing conclusions at lower and higher redshift than $z_{ref}$. The net outcome is that, within our framework, the HI $P(k)$ remains a high signal-to-noise measurement up to $z\simeq5$, a result that could be relevant for future radio surveys aiming to observe the 21 cm transition on cosmological scales.

\section{Acknowledgments}
We are happy to thank Uros Seljak, Andreu Font-Ribera and Pat McDonald for useful conversations, and Martin White for invaluable comments on the draft.

\bibliographystyle{mn2e}
\setlength{\bibhang}{2.0em}
\setlength\labelwidth{0.0em}
\bibliography{}

\end{document}